# Observation of new plasmons in the fractional quantum Hall effect: interplay of topological and nematic orders

Lingjie Du[1*], Ursula Wurstbauer[2,3], Ken W. West[4], Loren N. Pfeiffer[4], Saeed Fallahi[5], Geoff C. Gardner[5], Michael J. Manfra[5], Aron Pinczuk[1,6]

[1] *Department of Applied Physics and Applied Mathematics, Columbia University, New York, New York 10027, USA*

[2] *Walter Schottky Institut and Physik-Department, Technische Universität München, Am Coulombwall 4a, 85748 Garching, Germany*

[3] *Nanosystems Initiative Munich (NIM), Munich, Germany*

[4] *Department of Electrical Engineering, Princeton University, Princeton, New Jersey 08544, USA*

[5] *Department of Physics and Astronomy, and School of Materials Engineering, and School of Electrical and Computer Engineering, Purdue University, IN, US;*

[6] *Department of Physics, Columbia University, New York, New York 10027, USA*

\* e-mail: ld2751@columbia.edu

**Collective modes of exotic quantum fluids reveal underlying physical mechanisms responsible for emergent complex quantum ground states. We observe unexpected new collective modes in the fractional quantum Hall (FQH) regime: intra-Landau-level plasmons in the second Landau level measured by resonant inelastic light scattering. The plasmons herald rotational-symmetry-breaking phases in tilted magnetic fields and reveal long-range translational invariance in these phases. The fascinating dependence of plasmon features on filling factor provide new insights on interplays between topological quantum Hall order and nematic electronic liquid crystal phases. A marked intensity minimum in the plasmon spectrum at Landau level filling factor $v = 5/2$ strongly suggests that this paired state, which could support non-Abelian excitations, overwhelms competing nematic phases, unveiling the robustness of the 5/2 superfluid state for small tilt angles. At $v = 7/3$, a sharp and strong plasmon peak that links to emerging macroscopic coherence supports the proposed model of a FQH nematic state at this filling factor.**

The interplay between quantum electronic liquid crystal (QELC) phases and quantum liquids is a key topic of condensed matter physics, which is intensely studied in emergent electron quantum fluids in two dimensional (2D) systems [1–11] as well as in high-temperature superconductors [8,12,13]. QELC phases that break full rotational symmetry are distinguished by translational invariance, for example, the smectic phase has translational symmetry in only one spatial direction (along stripe-like structure) and the nematic phase is the one that retains long-range translational invariance. The partially filled second Landau level (SLL) can be regarded as a highly controllable venue for studies of interplays between QELC phases and quantum Hall fluids. More importantly, the quantum Hall fluids in the SLL include the enigmatic even-denominator fractional quantum Hall (FQH) state at $v = 5/2$, described as a *p*-wave paired superfluid phase [14–17], and new unconventional odd-denominator FQH states.

In the SLL, the interplays between topological order (in FQH phases) and QELC order (in nematics) have been mainly studied by low-frequency magneto-transport [2,11,18–20] that manifests the broken rotational symmetry of QELC phases via anisotropic transport. At $v = 7/3$, the application of even small in-plane magnetic fields, causes a marked transport anisotropy that coexists with the quantized Hall plateau [18]. The results indicate formation of emerging QELC phases that are closely linked to FQH edge states, leading to interpretations in terms of a new state of electron matter with FQH states that occur in the environment of a nematic phase (FQH nematic phase) [6,7,9,10,21–23]. However, transport anisotropies might come from anisotropic composite Fermions dressed by roton clouds [24], or a phase-separated two-dimensional electron system [18]. Recently, anisotropic transport with isotropic activation energies reported at $v = 5/2$ is interpreted as formation of a FQH nematic phase [20]. On the other hand, a compressible nematic phase could occur at $v = 5/2$, which takes over the 5/2 FQH state at sufficiently large in-plane magnetic fields [2,11,19,25,26]. Nematic phases have broken rotational invariance revealed by anisotropic transport experiments. However, long-range translational invariance, which could characterize nematic phases, is hard to access.

Here we report the observation of new collective modes in the partially populated SLL in tilted magnetic fields. The modes are identified as plasmon-like excitations from nematic phases that occur in the filling factor range $2 < v < 3$ of the SLL. Observed plasmon energies are well below the cyclotron energy, showing that the modes are intra-LL plasmons. To the best of our knowledge, such collective modes have not been reported or considered in quantum Hall systems. It is significant that intra-LL plasmon modes are observed at specific filling factors of FQH states as well as at non-FQH filling factors in a manner that offers new insights on the interplay between nematic and FQH orders that can now be directly probed in the *bulk*, without the intermediacy of edge states that occurs in transport.

The intra-LL plasmons in the SLL are observed by resonant inelastic light scattering (RILS) methods. The measured collective mode energy is described by

$$\omega(q) = (1+\xi)\omega_p(q), \qquad (1)$$

where $\xi \ll 1$ is a dimensionless parameter. $\omega_p(q)$ is the plasmon energy of a 2D electron system that has full translational symmetry [27],

$$\omega_p = \sqrt{n^* e^2 q/(2\varepsilon\varepsilon_0 m^*)}, \quad (2)$$

where $q = k$, and $k$ is the wave vector transferred in light scattering experiments. $m^*$ is the band mass of electrons in the GaAs quantum well. $\varepsilon$ and $\varepsilon_0$ are the background dielectric constant and free space permittivity. The parameter $n^*$ is a quasiparticle density in the spin-up SLL, which depends on magnetic field. The interpretation in terms of the plasmon energy of an electron system with full translational symmetry suggests that the plasmon wave vector $q$ is largely a good quantum number for wave lengths $\lambda_{//} = 2\pi/q$ much larger than characteristic periods in QELC phases. Then, the QELC phases that support the observed intra-LL plasmons preserve long-range translational invariance at the wavevector and temperature accessible in this experiment. Given that nematics are the QELC phases that have translational invariance, the modes described by Eqs. (1) and (2) are intra-LL plasmons in nematic phases in the partially populated SLL.

The observation of plasmons from nematic phases in partially populated LLs demonstrates a new venue to explore exotic quantum liquids in quantum Hall systems. At $v = 5/2$, the intensity of the plasmon line in the SLL is found to have a marked minimum. This is a result that uncovers a trend in which the even-denominator FQHE state at $v = 5/2$ overwhelms the nematic phase. The 5/2 state is currently understood as a superfluid gapped state of paired composite fermions. The minimum in the plasmon intensity suggests a suppression of the nematic phase that could be due to the tendency of a superfluid phase to extend to the whole 2D electron system. The robustness of the 5/2 state could be crucial in potential applications of the state in topological quantum computation [17].

Quite surprisingly, as we approach the filling factor $v = 7/3$, the intensity of a very sharp plasmon line is strongly enhanced. This unanticipated enhancement indicates that as topological FQH order appears at $v = 7/3$ there is *no* competition with the nematic order but rather that topological order and the nematic order coexist at $v = 7/3$. This observation is regarded as direct evidence for the proposed frameworks of nematic FQH states [6,7,9,10,21–23] and represents the application of an incisive experimental probe to the study of a novel quantum many-body phase that simultaneously displays both topological order and broken geometry symmetries.

The ultraclean 2D electron system is confined in 30-nm-wide, double-side modulation-doped GaAs/AlGaAs quantum well. The electron density is $n = 2.9 \times 10^{15}$ m$^{-2}$ and the mobility is $\mu = 23.9 \times 10^2$ m$^2$/Vs at 300 mK. The high sample quality enables RILS observations of several gapped FQHE states in the SLL [28–30]. The backscattering geometry described in Fig. 1A is employed in optics measurements with a small tilt angle $\theta \approx 20°$ in a dilution refrigerator operating at a base temperature below 40 mK. In RILS, there is a finite wave vector transfer $k = |k_L - k_S|\sin\theta = (2\omega_L/c) \sin\theta = 6.4 \times 10^2$ m$^{-1}$, where $k_L$ and $k_S$ are wave vectors of the incident and scattered photons, $\omega_L$ is the incident photon energy and $c$ is the speed of light in vacuum.

Figure 1B illustrates RILS spectra at filling factor $v = 7/3$ showing the resonance-enhanced mode at 1.43 meV. Similar modes exist in both FQH and non-FQH states, at energies that depend on filling factor in the range $2 < v < 3$, as shown in Fig. 1C. Nevertheless, the mode is absent at filling factor $v = 2$ and 3. The mode energies are significantly higher than those of neutral gap excitations of FQH states and of spin-wave excitations [28–30] below 0.15 meV (see Fig. 1D), but is much lower than those of inter-LL magnetoplasmons [31].

In the partially populated SLL, the quasiparticle density $n^*$ is equal to the degrees of freedom for intra-LL transitions in the partially populated SLL. For $v < 5/2$, $n^*$ is given by the electron density in the partially populated SLL $n_e = n - 2eB_\perp/h$, where $B_\perp$ is the perpendicular magnetic field and $h$ is the Planck constant. For $v > 5/2$, the Pauli exclusion principle limits the degrees of freedom to the density of holes (empty states in the SLL) given by $n_h = 3eB_\perp/h - n$. Indeed, Fig. 2A reports that for $2 < v \leq 5/2$ the plasmon energy is proportional to the square root of $n_e$ and for

$5/2 \leq v < 3$ the energy is proportional to the square-root of $n_h$. Figure 2B displays the mode energies in a single plot, revealing a slight difference between plasmons linked to $n_e$ and $n_h$. Well defined plasmon energies suggest wave vector conservation $q = k$ in RILS experiments. We fit the square-root dependences using Eqs. (1) and (2) with $k = 6.4 \times 10^4$ cm$^{-1}$, the value of $k$ at tilt angle $\theta = 20°$, and $m^* = 0.07\ m_0$ the band mass of electrons in GaAs. We treat $\xi$ as an adjustable parameter. The best fits shown in Fig. 2B as dashed lines are achieved with $\xi = 0.09$ for $v \leq 5/2$ and $\xi = 0.14$ for $v > 5/2$, respectively. As sketched in Fig. 2C, the transferred wave vector $k // B_{//}$ is perpendicular to stripelike structures in nematic phases [2,11]. A typical spacing between stripes $d$ is understood to be on a length scale of a few magnetic lengths $l \approx 11$nm [1,26,32,33], which is much shorter than the plasmon wavelength $\lambda_{//} = 2\pi/k = 0.98$ μm. Thus, plasmon excitations have long wavelength, with the local non-uniform charge density in nematic phases not resolved. The determined parameter $\xi << 1$ suggests that the measured points are described by a plasmon equation with a small plasmon wave vector $q=k$. For this reason, $q$ is a good quantum number in the long wave length limit ($\lambda_{//} >> d$), revealing the long-range translational invariance in the QELC phase.

Broken rotational invariance in emergent nematic phases in the partially populated SLL is revealed in RILS measurements of long wavelength spin-wave excitations shown in Fig. 1D. At $v = 3$ the electron system in the SLL has full in-plane rotational invariance and the spin wave occurs at the bare Zeeman energy as required by Larmor's theorem [28–30]. As the plasmon modes appear for $v < 3$, the spin wave modes soften due to broken rotational invariance of emergent nematic QELC phases and recover at the bare Zeeman energy for $v = 5/2$ [28].

The observations of plasmon modes of nematic phases offer direct insights on interplays of nematic order and the topological order in FQH liquids. Figure 3A shows normalized integrated intensities $I^* = I/n^*$, where $I$ is the integrated intensity of the plasmon peak, $n^* = n_e$ for $v < 5/2$ and $n^* = n_h$ for $v > 5/2$. The measured values of $I^*$ are close to 3 at filling factors away from FQH states. Surprisingly, at the $v = 7/3$ FQH state, $I^*$ increases by about a factor of two. This is a significant result that indicates remarkable links between the nematic phase and the FQH state at $v = 7/3$. In our interpretation we consider a model of plasmons in nematics in which the mode wave vector $q$ has real and imaginary parts Re $q$ and Im $q$. Wave vector conservation in RILS means that $k$ = Re $q$. A nonzero value of Im $q$ brings a finite plasmon coherence length $L = 2\pi$/Im $q$, which results in small breakdown of wave vector conservation selection rule in light scattering process [34]. As a result, RILS occurs in modes with a range $\Delta q$= Im $q$ =$2\pi/L$. In this model, the maximum RILS intensity of the plasmon line is proportional to $1/\Delta q^2 \propto L^2$. Thus the enhanced plasmon intensity at $v = 7/3$ demonstrates an enhanced value of $L$ as the FQH state at $v = 7/3$ is reached. The breakdown of exact wave vector conservation in RILS would result in a full width at half maximum (FWHM) $\Delta\omega=\omega(q)\Delta q/q$ of the plasmon line that is used to estimate $L$. Figure 3B displays that at $v = 7/3$ the length $L$ reaches its maximum value. These results could be linked to macroscopic coherence in FQH nematics when the emergent topological FQH order has long range correlations that favor translational invariance. We are thus led to propose that the dependence of plasmon intensity and lineshape in RILS spectra displayed in Fig. 3 gives crucial support to FQH nematic frameworks that describe the states near $v = 7/3$. Figures 3A also shows a plasmon intensity maximum at a filling factor close to $v = 8/3$, the particle-hole conjugate state of the 7/3 state, indicating a possible nematic presence in the FQH phase at $v = 8/3$. The smaller $L$ (larger $\Delta\omega$) for $v > 5/2$ indicates that hole-like plasmon excitations are more sensitive to disorder.

Figure 3B shows that there is continuity in the plasmon coherence length for filling factors in the vicinity of $v = 5/2$, which suggests that the plasmon mode at $v = 5/2$ is from a phase similar to that at non-FQH states, i. e. a compressible nematic phase, not the 5/2 FQH state. Figure 3A shows that at $v = 5/2$ the normalized integrated plasmon intensity collapses by nearly 80% from the intensity at non-FQH states (shown in the black dashed line in Fig. 3A). As illustrated in Figs. 4A and 4B, the plasmon intensity recovers for small changes $|\Delta v| \leq 0.01$ away from $v = 5/2$. The sudden drop of the plasmon intensity suggests that the nematic order is replaced by the FQH topological order in a quantum phase transition at $v = 5/2$. In other words the incompressible 5/2 state, proposed as the superfluid of paired composite fermions, dominates over the competing nematic phase. This interpretation is indeed consistent with the the recovery of rotational invariance manifested in the appearance of long wavelength spin waves at the Zeeman energy as shown in Fig. 1D for $v = 5/2$. The collapsing plasmon intensity confirms that the gapped 5/2 state built from composite fermions would not support a plasmon mode described by Eq. (1). The distinct behaviors of the plasmon modes at $v = 5/2$ and 7/3 here clearly reveal that the robust 5/2 state suppresses the nematic phase while the 7/3 state has a phase transition to the nematic FQH state.

In conclusion, observations of new intra-Landau-level plasmon modes of nematic phases provide new methods to explore the complex interplays between topological order and symmetry-breaking phases of 2D electron systems. The sharp and intense plasmon peak in the state at $v = 7/3$ reveals that the macroscopic coherence of the FQH liquid coexists with nematic order, which is key signature of a FQH nematic phase. In contrast, the collapse of the plasmon mode intensity at $v = 5/2$ uncovers that the robust gapped superfluid phase competes with and takes over the nematic phase.

**Acknowledgements**
The work at Columbia University was supported by the National Science Foundation, Division of Materials Research under award DMR-1306976. The Alexander von Humboldt foundation partially supported the experimental work at Columbia University. The MBE growth and transport characterization at Princeton was supported by the Gordon and Betty Moore Foundation through the EPiQS initiative Grant GBMF4420, and by the National Science Foundation MRSEC Grant DMR 1420541. The MBE growth and transport measurements at Purdue are supported by the U.S. Department of Energy, Office of Basic Energy Sciences, Division of Materials Sciences and Engineering under Award No. DE-SC0006671.

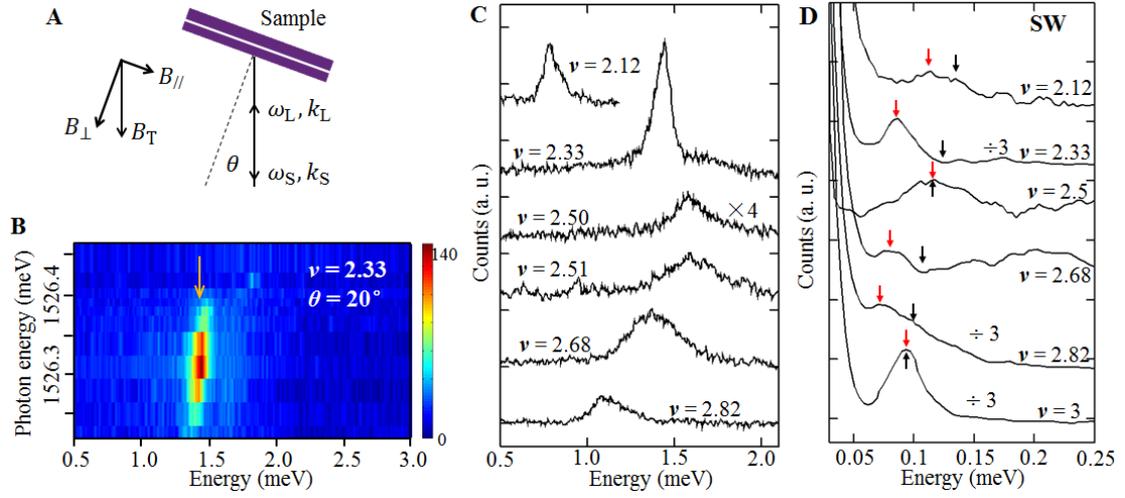

Fig. 1. (A) is a schematic description of the light scattering geometry at a tilt angle $\theta$. Incident and scattered light have photon energy $\omega_L$ and $\omega_S$, and wave vector $k_L$ and $k_S$. The total magnetic field $B_T$, the perpendicular component $B_\perp$ and inplane magnetic field $B_{//}$ are also shown. $k = 6.4 \times 10^4$ cm$^{-1}$ for $\theta = 20°$ and $\omega_L = 1526$ meV. (B) Color plot of resonant inelastic light scattering spectra of the plasmon measured at $\theta = 20°$ in the quantum electronic liquid crystal phase at $\nu = 2.33$ as a function of $\omega_L$. The mode intensity is resonantly enhanced at the plasmon energy (1.43 meV) marked by an arrow. (C) Resonant inelastic light scattering spectra of plasmons at filling factors in the range $2 < \nu < 3$. Intensities of the spectra in (B) and (C) are normalized to the incident light intensity. (D) Spin wave (SW) modes in the range $2 < \nu \leq 3$. Red arrows mark the position of the spin wave modes and black arrows indicate the Zeeman energy.

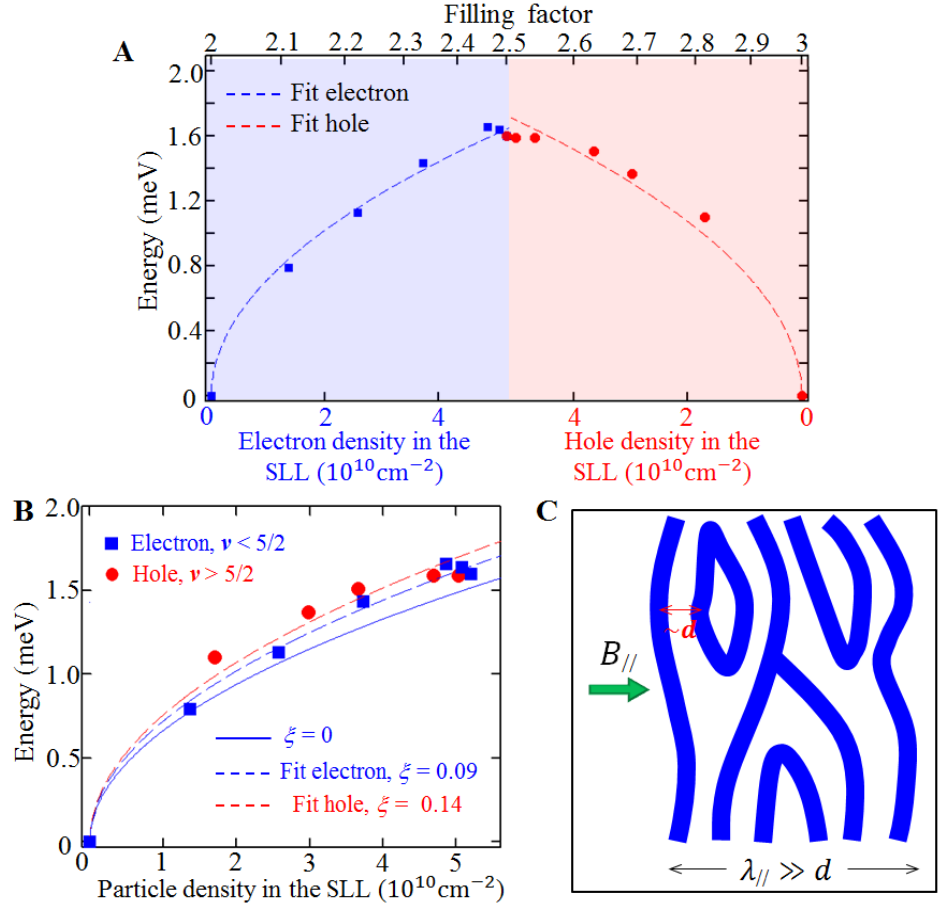

Fig. 2. (A) Plasmon energy as a function of particle density in the filling factor range of the spin-up second Landau level. The blue dots are for electrons and the red dots are for holes. The blue dashed line is a fit with a square-root dependence on electron density and the red dashed line is a fit with a square-root dependence on hole density. (B) Data points in (A) as a function of particle density $n^*$. The square-root dependence is described by Eq. (1) where $m^* = 0.07\ m_0$, $q = 6.4 \times 10^4$ cm$^{-1}$, $\varepsilon = (\varepsilon_{GaAs} + \varepsilon_{AlGaAs})/2 = 12.5$, $\varepsilon_{GaAs}$ is the dielectric constant of GaAs and $\varepsilon_{AlGaAs}$ is the dielectric constant of AlGaAs, α = 0.09 for electrons and α = 0.14 for holes. (C) Cartoon showing charge stripes in the second Landau level under an inplane magnetic field. For the wave vector $q$ // $B_{//}$ (see Fig. 1A), the plasmon wavelength $\lambda_{//} = 2\pi/q$ is much larger than a typical spacing between stripes (a few magnetic lengths).

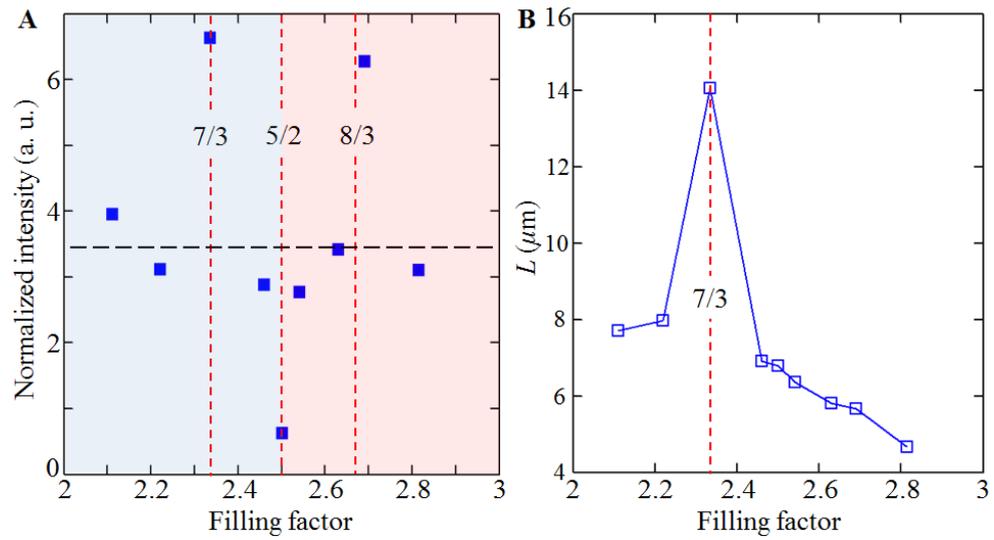

Fig. 3. (A) Normalized integrated intensity of the plasmon peaks as a function of filling factor. The integrated intensity is normalized by particle density in the second Landau level. The dashed black line marks a background value of electronic liquid crystal phases in the second Landau level. Three dashed red lines mark filling factors of $v$ = 7/3, 5/2 and 8/3. (B) Characteristic plasmon coherence length $L$ as a function of filling factor. $L$ is determined from the FWHM of the plasmon lines.

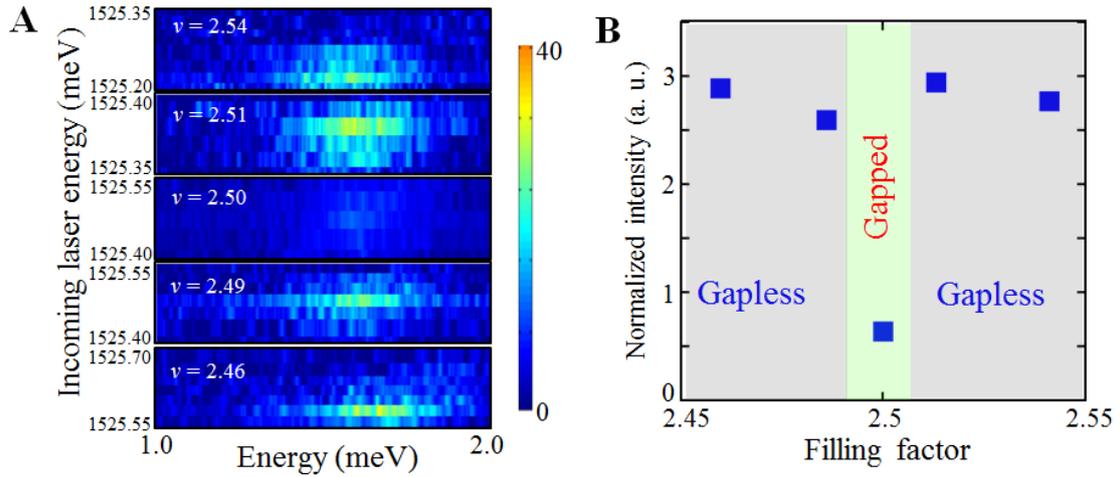

Fig. 4 (A) Color plots of resonant inelastic light scattering spectra measured at filling factors around $v = 2.50$ as function of incoming photon energy $\omega_L$. Resonance of the plasmon modes appears at close $\omega_L$ for filling factors around $v = 5/2$. (B) Normalized integrated intensity of the plasmon peaks around $v = 2.50$. The green area indicates the appearance of the gapped (incompressible) FQH state while the gray area indicates gapless (compressible) FQH states [27].